\documentclass[aps,onecolumn,prd]{revtex4}
\tighten

\input epsf
\usepackage{graphicx}
\newcommand{\beq}{\begin{equation}}
\newcommand{\beqa}{\begin{eqnarray}}
		  \newcommand{\eeq}{\end{equation}}
\newcommand{\eeqa}{\end{eqnarray}}

\newcommand{\lsim}{\lesssim}
\newcommand{\gsim}{\gtrsim}

\newcommand{\vect}[1]{\mbox{\boldmath${#1}$}}
\newcommand{\lmk}{\left(}
\newcommand{\rmk}{\right)}
\newcommand{\lnk}{\left\{ }
\newcommand{\rnk}{\right\} }
\newcommand{\lkk}{\left[}
\newcommand{\rkk}{\right]}

\newcommand{\p}{\partial}

\newcommand{\ver}{{\vect r}}

\newcommand{\vep}{{\vect p}}

\newcommand{\ves}{{\vect s}}

\newcommand{\veX}{{\vect X}}

\newcommand{\vR}{{\vect R}}

\newcommand{\dvp}{{\dot \varpi}}

\begin{document}

\title{ Relativistic Resonant Relations between Massive Black Hole Binary
 and Extreme Mass Ratio Inspiral } 
%
%
\author{Naoki Seto}
\affiliation{Department of Physics, Kyoto University
Kyoto 606-8502, Japan
}
\date{\today}

%
%
%
%
\begin{abstract}
One component of a massive black hole binary (MBHB) might capture a small third 
body, and then a hierarchical, inclined triple system would be formed. With the 
post-Newtonian approximation including  radiation reaction, we analyzed the  
 evolution of the triple initially with small eccentricities. 
We found that an essentially new resonant relation could  arise in the triple 
system.  Here relativistic effects are crucial. Relativistic resonances, 
including the new one, stably work even for an outer MBHB of comparable masses,   
and significantly change the orbit of the inner small body.

\end{abstract}
\pacs{PACS number(s): 95.85.Sz 95.30.Sf}
\maketitle

\section{Introduction}	
Merger of two massive black holes (MBHs) is one of the most violent events in 
the universe. A huge amount of energy is released in the form of gravitational 
waves (GWs). For a MBH binary (MBHB) with two masses $\sim 10^5$-$10^6M_\odot$, 
the Laser Interferometer Space Antenna \cite{lisa} can easily detect 
the waves from virtually anywhere in the universe. 
Meanwhile, GWs  from  an extreme-mass-ratio inspiral [EMRI, more  specifically, a compact 
object ({\it e.g.} neutron star, white dwarf) orbiting around a MBH]  would enable us to closely 
examine gravitational theories using a map of the spacetime around the MBH, encoded in the waves 
\cite{Ryan:1997hg,AmaroSeoane:2007aw}.  

In a MBHB, either of MBHs might trap a small third object and 
form an EMRI ({\it e.g.}  \cite{Chen:2010wy,Wegg:2010um,Yunes}).
Such a compound triple  system would be intriguing for cosmology and astrophysics. 
For example, 
with a tidal disruption of the small body, the MBHB merger could have precursive 
electromagnetic-wave signals 
that might allow us to identify the host galaxy of the MBHB and its redshift. 
Then we could observationally constrain the dark energy, using the luminosity 
distance estimated from the measurement of strong GWs emitted by the MBHB \cite{Schutz:1986gp}.

In the past $\sim 200$ years, orbital resonances have been ubiquitously identified among planetary 
or satellite systems obeying Newtonian dynamics \cite{ssd}. For example, the mutual stability of 
Pluto and Neptune is maintained by their 3:2 orbital periods around the Sun. 
Therefore, one might expect that orbital resonances can be an effectual mechanism 
 to form a stable three-body system including two MBHs. Indeed, 
except for detailed 
points, it was found that the resonant relations similar to those known in planetary/satellite 
systems  could set up relativistic triple systems  evolving by 
emitting gravitational radiation (see {\it e.g.} \cite{Seto:2011tu} for first 
order resonances such as 2:1 and also  \cite{Seto:2010rp} for  co-orbital ones). 
As in the case of Newtonian systems, the mass ratio of a MBHB involved in these  familiar and 
strong  resonances should be much smaller than unity 
(typically $\lsim O(10^{-2})$) \cite{Lee:2008nv}. However, from the viewpoint 
of gravitational wave astronomy, it is preferable that a MBHB of comparable 
masses can resonantly trap a small third body.

In this paper,  we report that an essentially new resonant relation could arise 
in a compound EMRI/MBHB system and it would significantly change the orbit of 
the inner EMRI. Here, relativistic effects, and inclined, hierarchical orbital 
configuration of the triple are crucial.   These  would be naturally 
realized for an EMRI/MBHB system, and the
 new resonance is among the two strong relativistic ones that work even for outer MBHBs 
of comparable masses, unlike the observed planetary or satellite systems.


This paper is organized as follows; in \S II we describe our numerical method 
and summarize basic notations. In \S III we  show a typical orbital evolution under 
the new resonant relation. \S IV is devoted to studies on the resonance. We 
first examine analytically how and why the resonance appears in our simulations, 
 and then provide results from systematic numerical analyses. We also  discuss issues 
 related to gravitational wave measurements. In \S V, we 
 roughly evaluate the expected capture rate of a small third body for a MBHB.  
 \S VI is a summary of this paper.

\section{Numerical Method}
In this paper, we numerically study  evolution of a triple system composed by a  MBHB 
(masses $m_0=(1-q)M$ and $m_1=q M$) and a small inner particle $m_2(\ll M)$ 
orbiting around $m_0$ (see Fig.1). Here $M$ is the total mass of the MBHB 
and $q$ is its mass ratio. We assume that $m_0$ and $m_1$ are comparable. 
The two objects $m_0$ and $m_2$ can be regarded as an EMRI  progressively 
perturbed by the outer body $m_1$ whose distance to $m_0$ shrinks due to 
gravitational radiation reaction.  Below, we adopt the geometrical unit 
with $G=c=M=1$, and do not include effects of spins.

For equations of motion of the triple, we use the  ADM Hamiltonian at 
2.5 post-Newtonian (PN) order (see {\it e.g.} \cite{Jaranowski:1996nv,Galaviz:2010te}) 
formally written by
\beq
H=H_N+H_1+H_2+H_{2.5}.
\eeq
Here $H_N$, $H_1$ and $H_2$ are the Newtonian, the 1PN and the 2PN terms 
respectively.    
$H_{2.5}$ is the first dissipative term induced by gravitational radiation.
The Hamiltonian $H$ is originally given for the position variables $\ver_{i}$ 
($i=1,2,3$: suffix for the particles) and 
their conjugate momenta $\vep_{i}$. But, instead of $\vep_{i}$, 
we introduce  new variables $\ves_{i}\equiv \vep_{i}/m_i$ and 
take  appropriate partial derivatives 
\beq
{\dot \ver}_{i}=\frac{1}{m_i}\frac{\p H}{\p \ves_{i}},~~
{\dot \ves}_{i}=-\frac{1}{m_i}\frac{\p H}{\p \ver_{i}}
\eeq
 in order to improve accuracy at numerically integrating  systems with  
 large mass ratios (the dot ${\dot {}}$ representing a time derivative). 
This prescription enables us to safely analyze the system even in the limit $m_2\to 0$.
But the three bodies $m_0,m_1$ and $m_2$ are handled equivalently in the post-Newtonian 
framework without introducing approximations associated with $m_2\ll m_0,m_1$. We numerically 
integrate the equations of motion using a fifth order Runge-Kutta method with an adaptive 
step size control (see \cite{Seto:2011tu} for details of the numerical method). 

We set up the initial conditions (denoted with the suffix ``$s$") of the compound 
EMRI/MBHB system, as follows. First, for the outer MBHB, we put its semimajor axis 
at $a_{1s}>300$   with the circular orbital velocity including the 1PN correction. 
We can realize a small initial eccentricity  $e_{1s}=O(10^{-4})$. 
Next, for the inner EMRI, we inject the small particle $m_2$ at the 
distance $a_{2s}(\ll a_{1s})$ from the moving body $m_0$, and set its relative 
velocity at the Newtonian circular velocity, to generate a small  eccentricity 
(typically $e_{2s}=O(10^{-2}))$.  
For simplicity, we mainly set the initial eccentricities $e_{1s}\ll 1$ and $e_{2s}\ll 1$. 
Actually, for a single MBH,  a majority of EMRIs might originate from dissolutions
of stellar mass binaries by the MBH \cite{ColemanMiller:2005rm}, and later 
become $e_{2}\ll 1$ due to radiation reaction. 
   In 
\S V we revisit issues related to the  orbital parameters of the inner EMRIs.
 Meanwhile, assuming independent 
evolutions of  EMRI/MBHB  in earlier epoch,
we randomly put their mutual orbital phase and  inclination.

To monitor the orbital elements, we define the coordinate distances between the 
three particles as
\beq
d_{1}\equiv d_{10}\equiv |\ver_1-\ver_0|,~~d_{2}\equiv 
d_{20}\equiv |\ver_2-\ver_0|,~~d_{12}\equiv |\ver_1-\ver_2|.
\eeq
 For $i=1,2$,  the semimajor axis $a_i$ and  eccentricity 
$e_i$ (including the forced ones)  are calculated from the  distance $d_i$ 
through its consecutive maximum ($a_i(1+e_i)$) and minimum ($a_i(1-e_i)\equiv r_{pi}$: 
pericenter distance). The MBHB evolves predominantly  by its gravitational radiation 
alone, and we have $a_1\simeq d_1$ with $e_1\sim e_{1s} (a_1/a_{1s})^{19/12}\ll 1$ 
\cite{Peters:1964zz}.
Since  we perturbatively include the relativistic effects, we  terminate our 
numerical integration when the distance $d_{ij}$ between any pair $i$-$j$ 
becomes less than 10 times $m_i+m_j$ (either C1: $d_1<10M$ 
or C2: $d_2<10m_0$ or C3: $d_{12}<10m_1$).

To define the angular variables of the inner and outer orbits, we introduce 
the Cartesian frame $XYZ$ around the central body $m_0$ with fixed spatial 
directions (see Fig.1).  Here the $XY$-plane is identical to the initial 
orbital plane of the MBHB. Because of  the initial inclination of the EMRI, 
the orbital plane of the MBHB slightly precesses. However, given  
$m_0,m_1\gg m_2$, the outer body $m_1$ virtually stays on the $XY$  
plane, and we use $\lambda_1$ for its angular position. 
Following standard conventions \cite{ssd}, we also define  $\Omega_2$ 
and $\varpi_2$ for the ascending node and the  pericenter of $m_2$. 
The inclination $I_2$ is the angle between the angular momentum of $m_2$ and the $Z$-axis.

To clarify the relative position between $m_1$ and $m_2$, we also introduce 
a complimentary frame $X_RY_RZ_R$ that is  corotating with $m_1$. The 
rotating  $X_R$-axis is oriented from $m_0$ toward $m_1$, and the $Z_R$-axis 
coincides with the original $Z$-axis. The coordinate values in the two frames are related by 
$\veX_R=\vR_Z(-\lambda_1) \veX$.   Here $\vR_Z(-\lambda_1)$ represents  the $3\times3$ 
rotation matrix around the $Z$-axis with the angle $-\lambda_1$ (and so on below).

\begin{figure}
  \begin{center}
\epsfxsize=6.5cm
\begin{minipage}{\epsfxsize} \epsffile{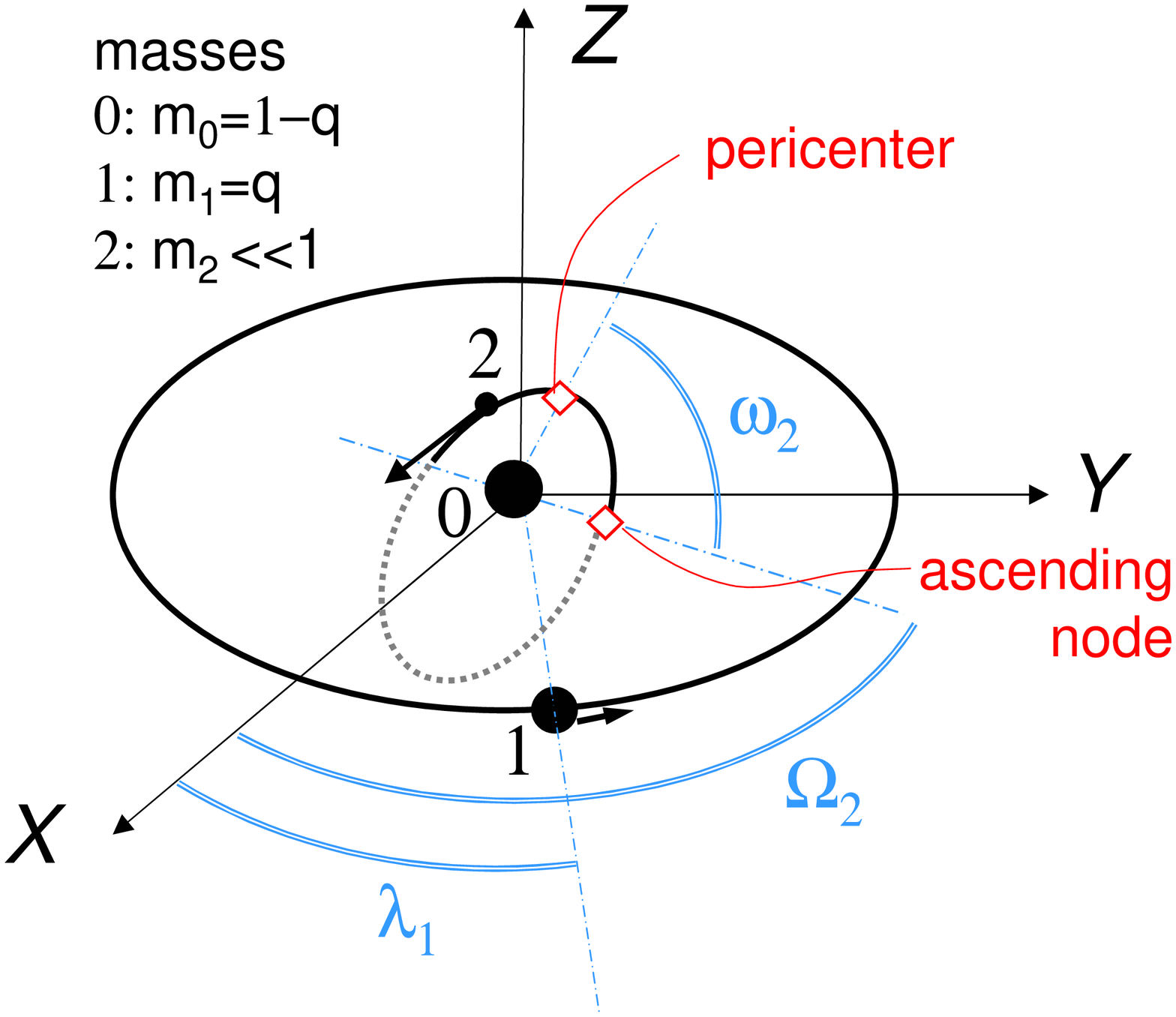}\end{minipage}
 \end{center}
  \caption{Configuration of the compound EMRI/MBHB system.  The $XYZ$-frame 
is defined around  the central body $m_0$, and its orientation is  spatially 
fixed. The massive outer body $m_1$  virtually remains on the $XY$-plane 
(its initial orbital plane and used as our reference plane)  with its 
angular position $\lambda_1$.  The inclined inner particle $m_2$  
intersects with the $XY$-plane  at the ascending node specified by $\Omega_2$. 
Its  pericenter is described by the angle $\varpi_2\equiv \Omega_2+\omega_2$ 
($\omega_2$: the angle between the ascending node and the pericenter). 
$\lambda_2$, $\Omega_1$ and $\varpi_1$ can be defined similarly, but less important in our study.
 }
\end{figure}

For reference, we provide useful expressions for a  binary with masses $m$ and $m'$, 
semimajor axis $a$, and the eccentricity $e$. The Kepler angular frequency $n$, 
the relativistic apsidal precession rate $\dvp_R$ by the 1PN term  \cite{LL1} are given by
\beq
n=\lnk \frac{m+m'}{a^3}\rnk^{1/2},~~\dvp_{R}=\frac{3 n (m+m')}{a(1-e^2)}.\label{base0}
\eeq
Meanwhile, the orbital parameters $a$ and $e$ decay  due to gravitational radiation 
reaction as \cite{Peters:1964zz}
\beqa
 \frac{-a}{\dot a}&=&\frac{5 a^4(1-e^2)^{7/2}}{64 (m+m') m m'} 
\lmk1+\frac{73e^2}{24}+\frac{37e^4}{96}  \rmk^{-1},\label{base1}\\
\frac{-e}{\dot e} &=&\frac{5 a^4(1-e^2)^{5/2}}{304 (m+m') m m'} 
\lmk1+\frac{121e^2}{304} \rmk^{-1}.\label{base2}
\eeqa

\section{Orbital Evolution}	
In Fig.2, we show  the orbital elements of the inner EMRI in one of our runs. 
The three masses are $m_0=0.6$, $m_1=0.4$ and $m_2=10^{-6}$.  We 
 set the initial orbital parameters, $a_{1s}=350$, $a_{2s}=37m_0$, $I_{2s}=74.3^\circ$, 
$e_{1s}\sim 10^{-4}$ and $e_{2s}\sim 3\times 10^{-3}$. In Fig.2, we use the outer 
semimajor axis $a_1$ as an effective time variable moving leftward from $a_1=350$ 
down to $\sim 160$.

At first, the EMRI evolves almost independently on the distant massive body $m_1$, 
and its orbital decay rate ${\dot a}_2$ is close to the analytical prediction (\ref{base1}).
Then, at $a_1\sim 290$, the eccentricity $e_2$ and the inclination $I_2$  start 
to increase rapidly. 
The orbit of the EMRI becomes retrograde at 
$a_1\lsim 240$.
Our  integration is ended at $a_1\sim 160$ by the  
condition C2 ($a_2(1-e_2)\simeq6$), when the time before the merger of the MBHB is $\sim5.4 (M/10^6M_\odot)$yr.

 In Fig.2d, we plot the combination of the angular parameters (modulo $2\pi$)
\beq
\varphi\equiv 3\lambda_1-\varpi_2-2\Omega_2 \label{30}
\eeq
sampled at intervals. This key  variable 
initially shows no structured pattern, but becomes localized around  
$\varphi\sim +0$ at $a_2\lsim 300$ (satisfying $|{\dvp}_2|\gg  |{\dot 
\Omega}_2|$). It is clear that the EMRI  is now resonantly trapped by the MBHB. 
To the best knowledge of the author, this is a new  resonant state never 
discussed in the literature. As we see later,  relativistic effects and the 
inclined, hierarchical orbital configuration are the crucial elements to raise 
the resonance. We also find that, for  $H=H_N$ (without the PN terms), the 
eccentricity $e_2$ of this highly inclined system is quickly increased by the 
Kozai process \cite{Kozai:1962zz} (easily destroyed by the relativistic apsidal motion  \cite{Holman}) 
and the run is ended shortly by the condition C2.

\begin{figure}[t]
  \begin{center}
\epsfxsize=9.5cm
\begin{minipage}{\epsfxsize} \epsffile{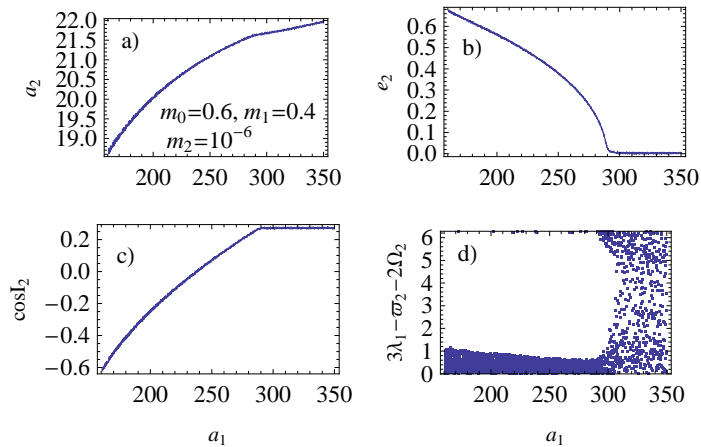}\end{minipage}
 \end{center}
  \caption{Evolution of the inner EMRI, as a function of the decaying  semimajor 
axis $a_1$ of the outer MBHB (leftward from $a_1=350$ down to 160). 
The panel (a) is for the inner semimajor axis $a_2$. The inner eccentricity $e_2$ 
(panel b) and the inclination $I_2$ (panel c) show sudden changes around $a_2\sim 290$ 
where the inner EMRI is resonantly captured by the outer MBHB, as shown in panel (d).
 }
\end{figure}

\begin{figure}
  \begin{center}
\epsfxsize=9.5cm
\begin{minipage}{\epsfxsize} \epsffile{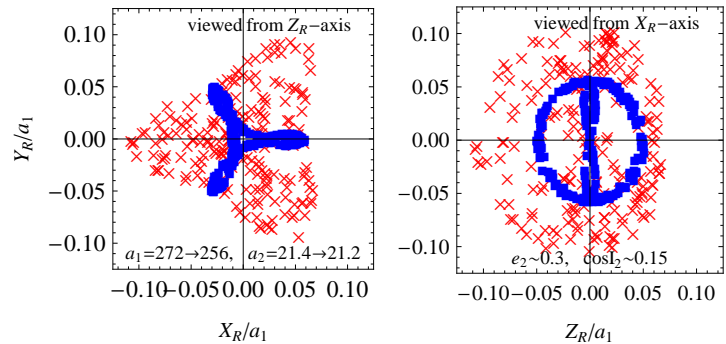}\end{minipage}
 \end{center}
  \caption{The positions (crosses) and pericenters (filled  squares) of the inner 
particle $m_2$ seen in the corotating $X_RY_RZ_R$-frame normalized by the outer 
distance $a_1$. In this frame, the central body $m_0$ is at  $(0,0,0)$ and the 
outer body $m_1$ is at $(1,0,0)$. The left panel is the projections of the 
points onto the $X_RY_R$-plane, and the right one onto the $Z_RY_R$-plane.   
These results are obtained from the same run as  Fig.2. The points are sampled 200 
times between $a_1=272$ and 256, corresponding to $\sim$3400 orbital cycles of the 
outer MBHB. The  pericenter of the EMRI is resonantly trapped by the MBHB, and 
stays nearly on the analytical curve (\ref{curve}) with $|\veX_R|=a_2(1-e_2)$.
 }
\end{figure}

The combination $\varphi$ indicates that the pericenter of the  inner particle 
$m_2$ has a simple geometrical relation to the position of the massive outer body $m_1$.  
In Fig.3, we plot the snapshots of the pericenter and the position of $m_2$ around the 
epoch $a_1\sim 265$. We use  the corotating frame $X_RY_RZ_R$, and the pericenter lies 
almost on a distinct one-dimensional structure. Indeed, the shape of the curve is 
roughly given by the simple analytical expression
\beq
\veX_R(u)
=a_2(1-e_2)\vR_Z(-u)\vR_X(I_{2})\lmk\begin{array}{@{\,}c@{\,}c@{\,}}
            \cos [3u]  \\
            \sin [3u] \\ 
	    0 \\
           \end{array} \rmk \label{curve}
\eeq
parameterized by
$u\sim\lambda_1-\Omega_2\sim(\varpi_2-\Omega_2)/3$.  Here the rotation $\vR_X(I_{2})$ 
represents the tilt due to the inclination and $\vR_Z(-u)$ is for the conversion to the 
corotating frame. 
The rapid evolution of $e_2$ and $I_2$ in Fig.2 can be understood as accumulation of 
coherent interactions between EMRI-MBHB though the established resonance.

\section{ Resonant modes}	
\subsection{Analytical Studies}

Now, with the aid of the disturbing function, we discuss why the resonant variable $\varphi$ 
appeared in our triple system. Roughly speaking, the disturbing function  is a perturbative 
expansion of the effective gravitational potential between $m_1$ and $m_2$ both orbiting 
around a central body $m_0$, and given  by a summation of terms proportional to
\beq 
\cos\lmk\sum_{i=1}^2 j_i\lambda_i+k_i\varpi_i+l_i \Omega_i  \rmk
\eeq
 with 
integers $j_i,k_i$ and $l_i$ \cite{ssd}.  The disturbing function is usually 
applied for systems with $m_0\gg m_1,m_2$ ({\it e.g.} planets around a star), 
but would  prove to be quite useful to interpret our numerical results with  $m_0\sim m_1\gg m_2$.

Since the outer MBHB continuously has  $e_1\ll 1$ and its orbital plane is almost unchanged, 
its pericenter and ascending node would not be important for the present resonant capture.  
Therefore, in the disturbing function,  we analyze the terms $A_{j_1j_2kl}\cos \phi_{j_1j_2kl}$  
with the phases
\beq
\phi_{j_1j_2kl}\equiv j_1 \lambda_1+j_2\lambda_2+k\varpi_2+l\Omega_2.
\eeq
 Below, we assume $j_1>0$, as we are interested in explicit resonant couplings between EMRI-MBHB.

We rely on the three basic properties generally valid for the individual terms of the 
disturbing function \cite{ssd,peale};  
\newline(i) the sum rule; $j_1+j_2+k+l=0$ from the rotational symmetry around the $Z$-axis, 
\newline(ii) the scaling relation of the amplitude; $A_{j_1j_2kl}=O\lkk e_2^{|k|} 
I_2^{|l|}(a_2/a_1)^p\rkk $  for $e_2,I_2,(a_2/a_1)\ll 1$ with $p\ge 2$, 
\newline(iii) the restriction of $l$ to even numbers: $l=2\nu$ ($\nu$: integer) from the 
symmetry with respect to the $XY$-plane.

A resonant state is identified by the condition $\phi_{j_1j_2kl}\simeq const$. 
Here we analyze a more tractable form ${\dot \phi}_{j_1j_2kl}\sim 0$, and 
evaluate the magnitudes $({\dot \lambda}_1,{\dot \lambda}_2,{\dot 
\varpi}_2,{\dot \Omega}_2)$ and the adequate integers $(j_1,j_2,k,l)$ for the  
triple system with the hierarchy $a_2\ll a_1$ and a small initial 
value $e_{2s}\ \ll1$. 

Because of the correspondence ${\dot \lambda}_i\sim n_i\propto a_i^{-3/2}$ (see eq.(\ref{base0})) 
and the general relation ${\dot \lambda}_i\gg {\dot \varpi}_i,  {\dot \Omega}_i$ valid also 
in weak field regime, our system satisfies
$
{\dot \lambda}_2 \gg  {\dot \lambda}_1, \dvp_2,{\dot \Omega}_2.
$
Then,  for $j_2\ne 0$, the condition ${\dot \phi}_{j_1j_2kl}\sim 0$ implies a very high-order 
resonance  with $|j_1+j_2|\sim |j_1|\sim |j_2| (a_1/a_2)^{3/2}\gg 1$.  Instead, 
we limit our analysis only for the simpler cases with $j_2=0$.

Next we compare $\dvp_2$ and ${\dot \Omega}_2$, by separately evaluating the secular
Newtonian effects from the distant body $m_1$ and the relativistic 
effects around the nearby one $m_0$. The Newtonian contributions 
$\dvp_{N2}$ and $ {\dot \Omega}_{N2}$ become $O(n_2(a_2/a_1)^{3})$ \cite{ssd}, 
while the 1PN ones are  $\dvp_{R2}\sim 3n_2(1-q)/a_2$ (see eq.(\ref{base0})) and  
${\dot \Omega}_{R2}=0$ \footnote{If the MBHs are spinning,  
$\Omega_2$ precesses {\it e.g.} due to the spin-orbit coupling at 1.5PN order. 
But, considering the self-adapting nature of orbital resonances, the spin and 
also other 
higher PN effects would not qualitatively change our results  at least 
for slowly rotating MBHs.}.
The total relation becomes $\dvp_{2}\simeq\dvp_{R2} \gg |{\dot \Omega}_{2}|\sim  |{\dot \Omega}_{N2}|$, 
consistent with our numerical results.
 Thus, for our weak field system, the resonance ${\varphi}_{j_1j_2kl}\simeq 
 const$ $(j_1>0)$ should be realized with the relation $j_1{\dot 
 \lambda}_1+k\dvp_2\sim0$ ($k< 0$). Here, a smaller $|k|$ is preferred from 
the scaling relation (ii), and we put $k=-1$.  In contrast, a small 
 inclination angle $|I_2|\ll 1$ is not assumed, and we do not need 
to impose a 
 strong requirement on  $l=2\nu$ at present. Then, the resonance 
variable valid for our system is written as
\beq
\theta_{;\nu}\equiv (2\nu+1) \lambda_1-\varpi_2-2\nu\Omega_2.
\eeq
Below, we attach ``;'' before the mode-number suffixes $\nu$  to distinguish
them from the labels $i$ for the particles.

As seen so far, the relativistic effects  are crucial to make ${\dot 
\theta}_{;\nu}\sim 0$ by  increasing the apsidal precession rate $\dvp_2$. In 
addition, the overall dissipative evolution is due to the 2.5PN radiation 
reaction force. At the resonance, the PN order parameter 
$(1-q)/a_2=O(\dvp_2/{n}_2) $ of the EMRI is  comparable to the orbital 
hierarchy 
$ (a_2/a_1)^{3/2}\sim { n}_1/{ n}_2$.   Thus, our sequence $\theta_{;\nu}$ 
contains only the outer position $\lambda_1$ without the inner one $\lambda_2$,  
remarkably different from the standard mean motion resonances with $j_1j_2\ne 0$. 
Here, the unusually large outer mass $m_1(\sim m_0$) would  enhance the resonant coupling.

The observed combination $\varphi$ in eq.(\ref{30}) is properly reproduced 
as $\varphi=\theta_{;1}$. Interestingly, for $\nu=0$, the variable  
$\theta_{;0}$ coincides with  the $j=1$ inner Lindblad resonance. 
The Lindblad resonances [here parameterized by 
$j\lambda_1-(j-1)\lambda_2-\varpi_2$] play fundamental roles 
in the dynamics of coplanar disks \cite{GD}
 whose relativistic effects are recently discussed in \cite{Hirata:2010vp}.

From a basic relation of the disturbing function ({\it e.g.} eq.(12) in 
\cite{peale}), we get $p=\max\{3,2\nu+1\}$ for the relation
(ii) with  the variables $\theta_{;\nu}$. This shows  weaker coupling for $\nu\ge 
2$.  Actually, even for $|k|=2$, we 
have $p\le 3$ only for $\phi=2\lambda_1-2\varpi_2$ that completely degenerates 
with $\theta_{;0}=\lambda_1-\varpi_2$.

In the C ring of Saturn, there is a ringlet structure whose longitude of the 
pericenter $\varpi$ is in a resonant relation with the angular position $\lambda_{T}$ of Titan, the 
largest satellite of Saturn. The resonant variable is given by 
$\lambda_{T}-\varpi$, and   the apsidal precession $\dvp$ of the ringlet is mainly
caused by  the large multipole 
moments of Saturn ({\it e.g.} its quadrupole moment $J_{20}=0.016$) \cite{porco}, 
instead of relativistic corrections.

Now we can predict when the inner  EMRI with $e_2\sim 0$ is resonantly 
captured by the outer MBHB. From the relation $(2\nu+1){\dot 
 \lambda}_1-\dvp_2\simeq0$ (or equivalently 
$2\nu+1\simeq{\dvp_{2R}}/{n_1}$), the critical inner semimajor axis $\gamma_{;\nu}$  is 
given as a function of $a_1$ by
\beq
\gamma_{;\nu}=3^{2/5} \lmk {2\nu+1}  \rmk^{-2/5} (1-q)^{3/5}  a_1^{3/5}. \label{rr}
\eeq
For $\nu=0$ and the test particle limit $q=m_1\to 0$, this expression 
coincides with eq.(130) in \cite{Hirata:2010vp} for the 
$j=1$ inner Lindblad resonance. 
Using the results in Fig.2, we also confirmed  that the ratio $\dvp_2/n_1\propto 
a_1^{3/2} a_2^{-5/2}/(1-e_2^2)$  is nearly constant during the trapping.

\begin{figure}
  \begin{center}
\epsfxsize=9.cm
\begin{minipage}{\epsfxsize} \epsffile{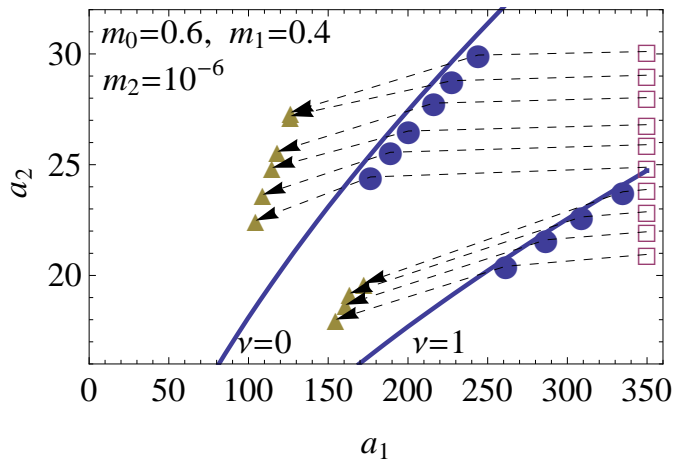}\end{minipage}
 \end{center}
  \caption{Evolution of the triple systems in the  $a_1a_2$-plane.  The open 
squares are the initial conditions with $a_{1s}=350$ and  $a_{2s}=20$-30. 
The  circles represent the points when $e_2$ exceeded 0.1 (indicating resonant 
capture), and the triangles are the termination points of the runs 
(all by the condition C2). The two solid curves are the analytical 
predictions (\ref{rr}) for the onsets of the resonant capture for $\nu=1$ and 0.
 } 
\end{figure}

\subsection{Numerical Studies}

Next we systematically analyze a series of numerical simulations. 
We take the mass parameters; $q=0.4$, $m_2=10^{-6}$ and the initial 
conditions; $e_{1s}< 10^{-4}$, $e_{2s}<10^{-2}$, $a_{1s}=350$.  
We prepared totally 11 runs from various initial inner separations 
$a_{2s}\in[33.3m_0,50m_0]$ with random  initial inclinations $\cos I_{2s}\in [0,1]$. 
For these sets $(a_{1s},a_{2s})$,  the infall rate is larger for the MBHB 
(namely $d(a_2/a_1)dt>0$).  From eq.(\ref{base1}), this 
catch-up condition is written as $a_2> \eta_c a_1$ with 
$\eta_c\equiv(m_0m_2/m_1M)^{1/4}$. When
we reverse the time, each non-resonant triple   moves on the curve
\beq
a_2^4-(\eta_c a_1)^4=const
\eeq
 and  asymptotically approaches to the line 
$a_2=\eta_c a_{1}$.

In Fig.4, we provide  the time evolutions of the runs. We obtained similar results 
without the 2PN term.
The run from $a_{2s}\simeq 22$ is what was already shown in Fig.2.  In 
Fig.4 we  added the analytical predictions $a_2=\gamma_{;\nu}$ ($\nu=0,1$)  for the onset of the 
resonant capture.  They reasonably agree with the numerical results.  Since 
 the initial outer distance $a_{1s}=350$ is not sufficiently 
large for $a_{2s}\gsim 25$,  the corresponding EMRIs are captured by 
the  $\nu=0$ mode. 
We also examined dependence of the resonant captures on the mutual inclination of the two 
 orbits.
It was  found that slightly inclined EMRIs ({\it e.g.} $I_{2s}=0.14$ and 0) could pass through the $\nu=1$ mode and first reacted to the $\nu=0$ mode, in accord with the scaling relation $A_{j_1j_2kl}\propto (I_2)^{2\nu}$.

We briefly describe other interesting results. First, to realize a capture, the 
resonant curve $a_2=\gamma_{;\nu}$ ($\nu=1,0$) in Fig.4 should 
be encountered in the direction ${\dot \nu}<0$.  
For  ${\dot \nu}>0$, 
 both $e_2$ and $I_2$ show gaps at 
the resonant crossings, but 
the capture was unsuccessful.
Using eq.(\ref{rr}) in the form $(2\nu+1)\propto a_1^{3/2}a_2^{-5/2}$, we can assign 
contour levels $\nu$ on the $(a_1,a_2)$ plane.  Then, from eq.(\ref{base1}), 
we obtain
\beq
{\rm sign}({\dot \nu})={\rm sign} \lmk \frac35 \frac{{\dot a}_1}{a_1}- 
\frac{{\dot a}_2}{a_2}  \rmk={\rm sign}(\kappa \eta_c 
a_{1}-a_2)
\eeq
with $\kappa\equiv (5/3)^{1/4}$. Therefore, we have ${\dot \nu}<0$ (required for captures) at $a_2>\kappa \eta_c 
a_{1}$.  In Fig.5, we provide a schematic illustration for the resonant capture.
We denote the intersection of  the line $a_2=\kappa \eta_c 
a_{1}$ with the resonant curve $a_2=\gamma_{;\nu}$   by 
$(x_{1;\nu},x_{2;\nu})$. For 
a capture by the $\nu$ mode, a triple should cross its resonant
curve at $a_1<x_{1;\nu}$ (the solid part in Fig.5). 
  Furthermore, as understood from the  flows of triples in the $(a_1,a_2)$-plane, 
  an  EMRI capturable either by $\nu=0$ or 1 must exist in 
$\kappa^{-1}x_{2;0}<a_2< x_{2;0}$ (shown by the double line in Fig.5) at the critical epoch $a_1=x_{1;0}$.

Secondly, 
for the parameters shown  in Fig.2, we examined  systems with larger 
 $e_{2s}$.  For $e_{2}\gsim 0.11$, the system transversed the $\nu=1$ curve
 without a capture but again showed gaps of $e_2$ and $I_2$. In the same manner, we 
 analyzed the crossing of the $\nu=0$ mode around $a_1\simeq 200$ and $a_2\simeq 
 29$ with various eccentricities $e_2$, and found an upper limit $e_2\sim 0.35$ 
 for yielding  captures.
 Similar behaviours ($e_2$- or ${\rm sign}({\dot 
\nu})$-dependence of the captures, observed gaps of $e_2$ at the resonant 
crossings without captures) are 
 found for standard mean motion resonances and well explained by the 
 separatrix structure  in an effective phase space (see Figs.3-6 in \cite{peale} 
 and also \cite{capture}).

By dropping the time consuming 2PN term, we also performed  runs from 
$a_{1s}\sim 600$ and $a_{2s}\sim30$ (corresponding to $\nu>3$ 
in Fig.4). But the captures by the $\nu>1$ 
modes did not occur, as anticipated from the arguments on the power $p$ in the 
relation (ii).  Furthermore, by 
artificially  multiply a large factor to the dissipative term $H_{2.5}$, we 
evolved 
widely separated systems ($a_1\sim2000$ and $a_2\gsim \gamma_{;0}\sim 100$ for 
$\nu=0$) at accelerated rates, and 
 confirmed captures by the $\nu=0$ mode.

\begin{figure}
  \begin{center}
\epsfxsize=8.5cm
\begin{minipage}{\epsfxsize} \epsffile{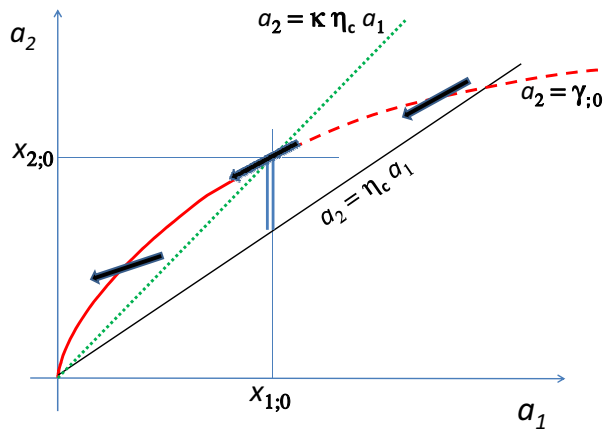}\end{minipage}
 \end{center}
  \caption{Resonant capture of the inner EMRI by the outer MBHB. We show a 
schematic illustration for the $\nu=0$ mode with a small eccentricity $e_2\simeq 0$. The thick (solid and dashed) curve 
is the critical  semimajor axis $a_2=\gamma_{;0}$ of the EMRI  for the resonant 
condition (\ref{rr}). The solid line $a_2=\eta_c a_1$ is the catch-up line where we have 
$a_1^{-1} da_1/dt=a_2^{-1} da_2/dt$.  The dotted line  $a_2=\kappa \eta_c a_1$ 
with $\kappa=(5/3)^{1/4}$
divides the signature of the resonance approach ${\rm sign({\dot \nu})}$, as 
shown by the three arrows. We put the intersection of the two curves by $(x_{1;0}, x_{2;0})$.
To realize a capture, the resonant curve  
$a_2=\gamma_{;0}$ should be crossed in the direction ${\dot \nu}<0$, 
corresponding to the upper region of the dotted line (the solid part of the curve $a_2=\gamma_{;0}$).
 Therefore,  at the
critical outer distance $a_1=x_{1; 0}$,  the capturable inner EMRI must has a 
semimajor axis $\kappa^{-1}x_{2;0}<a_2< x_{2;0}$ shown by the vertical double line. }
\end{figure}

\subsection{Gravitational Wave Measurements}

Here, we discuss GW detection for a trapped EMRI. In Fig.2, the final infall 
rate ${\dot r}_{p2}=d[a_2(1-e_2)]/dt$ is $\sim 70$ times larger than that of a corresponding isolated 
EMRI (see eqs.(\ref{base1}) and (\ref{base2})).
 Considering the effective time 
duration of the GW signals,  the detectable distance 
for the trapped one becomes $\sim1/\sqrt{70}$ times smaller. This is partly due 
to 
the small 
inner mass $m_2=10^{-6}$, and, indeed, the results in Fig.2 are similar to those 
for a test particle with $m_2=0$. We thus analyzed systems with larger $m_2$ from 
$(a_{1s},a_{2s})=(305,21.5)$ now down to $r_{p2}=6m_0$. The resonant 
trapping was successful up to $m_2\sim 10^{-5}$  for which the trapping  
ended at $r_{p2}\sim 8m_0$ with the final rate  ${\dot r}_{p2}$ close to the 
isolated EMRI. But  the corrections higher than 2.5PN could become important 
here.

To detect a trapped EMRI, a simple data analysis worth trying is a  search triggered by strong GW 
of a merging MBHB that provides {\it e.g.} $m_1$, $m_2$, the sky location of the system and also $a_1$ as a function of time. 
Here the primary orbital parameters of the EMRI would be its semimajor axis $a_2$, 
eccentricity $e_2$ and mutual inclination $I_2$.  These should be searched by 
using templates.  The trapping condition $a_1^{3/2}\propto a_2^{5/2}(1-e_2^2)$
might become useful to  narrow  the parameter space of   $a_2$ and $e_2$ to be surveyed.

Now we  go back to the specific  system shown in Fig.2, as an example, and 
discuss its final phase around $a_1\sim 160$ and $a_2\sim 19$.  The orbital periods of the EMRI and the 
MBHB are $\sim 700$ and $\sim 13000$ respectively. Since the EMRI has a large 
eccentricity, it emits relatively strong pulse-like GWs around its pericenter.
These GWs would be an interesting observational target.
The characteristic duration of each pulse is $(r_{p2}^3/m_0)^{1/2}\sim 20$ and 
the interval between the adjacent pulses  is approximately the orbital period of the EMRI.  Therefore, the 
semimajor axis $a_2$ is the critical parameter for matching the GW signals.

For a standard mean motion resonance, the inner and outer semimajor axes are 
related through a simple  linear relation between the two orbital periods.  For example, we 
 have $a_2/a_1\simeq [j/(j+1)]^{2/3}$ for the first order resonance with the librating 
 variable $j\lambda_2-(j+1)\lambda_1\sim const$ ($j\ge 1$: an integer). This kind of relation  would be quite 
 helpful to estimate evolution of the inner axis $a_2$ from the observed GW signals of the 
 merged MBHB.  However, our hierarchical resonances $\theta_{;\nu}$ do not 
 explicitly depend on the angular position $\lambda_2$ of the trapped  EMRI, and 
 we cannot directly deduce the key parameter $a_2$ for the EMRI,  as a function 
 of $a_1$. Therefore, for an EMRI trapped in  our resonances,   the 
 identification of its GW signature  could become more demanding, compared with 
  standard mean motion resonances.

\section{ capture rate}	

In this section, we roughly discuss probability of a MBHB  ($2\times 
10^6M_\odot+10^6M_\odot$) trapping a $10M_\odot$ BH at the $\nu=0$ and 1 modes.  
With respect to these mass parameters, the critical point in Fig.5 is given by
\beq
(x_{1;0},x_{2;0})=(2040,118).
\eeq

The formation scenario of an  EMRI around a single MBH (not around 
a MBHB)  often studied in the literature is the capture 
of a compact object around its close approach 
to the MBH  by emission of gravitational radiation (hereafter two-body capture 
process) \cite{Freitag:2002nm,Hopman:2005vr,Merritt:2011ve}.  However, the formed EMRI would 
be highly eccentric, 
and the  residual eccentricity has a typical value $e_2\sim 0.5$ even when 
entering into  the LISA band, 
 corresponding to $a_2=O(10)$ \cite{Hopman:2005vr}. Therefore, considering the 
 thresholds for the  
inner eccentricity 
$e_2$ discussed in \S IV.B, the resonant capture by an outer MBHB would not be promising for 
the EMRIs formed by the two-body capture process.

Miller et al. \cite{ColemanMiller:2005rm} discussed formation of an EMRI through 
a tidal dissolution of a compact stellar binary by an MBH (hereafter binary
dissolution process).  One component of the binary is expelled from the system 
as a high-velocity star \cite{Yu:2003hj} and another one becomes bound to the MBH.  
 The resultant 
EMRI  
could have a larger initial pericenter distance (thus a larger cross section for 
 its formation), 
compared with the  two-body capture process, and could has comparable 
or larger population.
 In addition, for the binary dissolution process, the residual eccentricity $e_2$ 
 becomes small $e_2\sim 0$ in the LISA band, due to a long-term dissipative 
 evolution of the orbit with emitting gravitational radiation.

Here, following \cite{ColemanMiller:2005rm}, we evaluate the orbital parameters of the inner EMRI formed by the binary
dissolution process, simply assuming that  the EMRI is affected by the secondary MBH 
solely through the $\nu=0$ and 1 resonances.

We first consider only the  evolution of the EMRI without taking into account 
the secondary MBH $m_1$. A compact 
stellar binary with an orbital separation $a_{bin}$ and mass $\sim m_2$ is dissolved by the 
primary MBH $m_0$ at the distance
\beq
d_{2,tide}\sim \lmk \frac{3m_0}{m_2}\rmk^{1/3} a_{bin}\sim 8.8 \lmk \frac{m_0}{2\times 10^6 
M_\odot}\rmk^{1/3}\lmk \frac{m_2}{10M_\odot}\rmk ^{-1/3}\lmk \frac{a_{bin}}{0.1{\rm AU}}\rmk{\rm AU},
\eeq
where we put the fiducial parameter $a_{bin}$=0.1AU given in 
\cite{ColemanMiller:2005rm}. We regard $d_{2,tide}$ as the initial pericenter 
distance of the EMRI and also use the typical value $e_2\sim 0.98$  quoted in 
\cite{ColemanMiller:2005rm} for the 
initial eccentricity. 

Then, with eqs.(\ref{base1}) and (\ref{base2}), the merger time of the EMRI is 
given by $\sim 5\times 10^9$yr (less than the age of the universe).  At 
$a_2=x_{2;0}=118$, we have   a significantly reduced eccentricity  $e_2\sim 
0.17$ and the 
remaining time $1.1\times 10^6{\rm yr}~(\propto a_2^4)$.  When the outer MBH inspirals down to 
the critical separation
$a_1=x_{1;0}=2040$, the distribution of the inner EMRIs at $a_2\le x_{2;0}$ 
would  be
in a steady state, from our assumptions. We can
characterize the distribution of the EMRIs by their infall rate  at $R_{in}=5\times 10^{-8}{\rm yr^{-1}}$  \cite{Freitag:2002nm}. This 
rate  $R_{in}$ was originally given for the two-body capture process, but we use 
it for the binary dissolution process, following the arguments in \cite{ColemanMiller:2005rm}.  Finally, the 
probability of a MBHB merger with a trapped $10M_\odot$ BH can be evaluated as 
\beq
P=(5\times 10^{-8}{\rm yr^{-1}})(1.1\times 10^6{\rm yr})(1-3/5)=0.022,
\eeq
corresponding to the expected number of EMRIs on the double line in Fig.5.

Actually, the above probability $P$ contains the contribution of the EMRIs that would 
be captured by 
the $\nu=1$ mode.  These EMRIs pass close to the lower bound 
$a_2=\kappa^{-1}a_{1;0}=103.9$ on the double line.
Since we have $x_{2;\nu}\propto 1/(2\nu+1)$, the branching ratio of these modes 
is given by 
\beq
1-3^{-4}:3^{-4}=80:1,
\eeq
and dominated by the $\nu=0$ mode. For inner EMRIs with white dwarfs or neutron 
stars of $m_2\sim O(1M_\odot)$, the merger times become larger than the age of 
the universe for the previous input parameters $a_{bin}=0.1$AU and $e_2=0.98$.  
But,   
if we can assume  small evolved eccentricities $e_2\lsim 0.2$ at $a_2= x_{2;0}$ and the
 steady-state distributions   normalized by $R_{in}=5\times 10^{-8}{\rm 
 yr^{-1}}$, the probabilities become $P=O(1)$ for these EMRIs.

 Thus far, we have considered only an isolated  three-body system in a simplified 
 manner. In reality, a pre-existing EMRI might be destroyed  by 
other stars ({\it e.g.} scattered from the secondary MBH). But, 
at the same time, the secondary MBH could highly enhance the capture rate of 
EMRIs $R_{in}$,  at least, for those  formed by the two-body capture process \cite{Chen:2010wy}. This is an interesting possibility and would be worth examined 
in the context of the binary dissolution process.

In ongoing or planned wide-filed surveys for transient electromagnetic waves (see 
\cite{Bloom:2011xk} for recent results), we might detect a tidal disruption event 
resonantly driven by a merging MBHB.
For such a event, the signature of the orbital period of the MBHB might be 
found in the temporal structure of the emitted electromagnetic waves (see 
\cite{Liu:2009kh} for related discussions).
 Here, an MBH more massive than $\sim 10^6M_\odot$ 
can directly swallow a white dwarf without a tidal  disruption, and we need to consider 
a main sequence star in this mass regime.

\section{ Summary}	

Orbital resonances might be a potential mechanism to append a small body to an 
inspiraling MBHB, and  to maintain the compound EMRI/MBHB system.  Based on the post-Newtonian approximation, we numerically examined 
such hierarchical three body systems evolved by emitting gravitational radiation.
For mutually inclined orbital configurations, we found a new resonant state with 
the librating variable $\theta_{;1}=3\lambda_1-\varpi_2-2\Omega_2$.  This 
resonant state, together with another state  $\theta_{;0}=\lambda_1-\varpi_2$, 
works relatively strongly for the triple systems.  Here the relativistic apsidal 
precession is essentially important, and the post-Newtonian parameter at the capture becomes 
comparable to the hierarchy of the two orbits $(a_2/a_1)^{3/2}$. In contrast to 
standard mean motion resonances known among planetary/satellite systems, these 
relativistic resonances can prevail even for outer MBHBs of comparable masses.
During these resonances, the eccentricity and inclination of the 
inner EMRI increase almost monotonically, and its pericenter distance could go down  
blow $\sim5$ Schwarzschild radii.

In order to realize a capture into the resonances, an inner  EMRI cannot have a  
large eccentricity.  Therefore, EMRIs formed through the familiar two-body capture 
process  
would be difficult to be involved in the resonances.  In contrast,      
 the binary dissolution process can produce mildly eccentric EMRIs at the 
  spatial scale in interest, and these EMRIs might be resonantly trapped by 
   inspiraling outer MBHBs.

The author would like to thank Takahiro Tanaka, Xian Chen and  anonymous referees for 
helpful comments.
This work was supported by JSPS grant 2074015.

\end{document}